\documentclass[11pt, a4paper]{article}
\usepackage{}

\usepackage{indentfirst}
\usepackage{lineno}
\usepackage{graphicx}
\usepackage{ae}
\usepackage{amsmath}
\usepackage{amssymb}
\usepackage{latexsym}
\usepackage{url}
\usepackage{epsfig}
\usepackage{cite}
\usepackage{mathrsfs}
\usepackage{amsfonts}
\usepackage{amsthm}
\usepackage{float}
\usepackage{booktabs}
\usepackage{subfigure}

\usepackage{color}

\long\def\delete#1{}

\newcommand{\be}{\begin{equation}}
\newcommand{\ee}{\end{equation}}
\newcommand{\ben}{\begin{equation*}}
\newcommand{\een}{\end{equation*}}
\newcommand{\bea}{\begin{eqnarray}}
\newcommand{\eea}{\end{eqnarray}}
\newcommand{\bean}{\begin{eqnarray*}}
\newcommand{\eean}{\end{eqnarray*}}

\numberwithin{equation}{section}

\title{Ranking the spreading influence of nodes in complex networks based on mixing degree centrality and local structure \thanks{Supported by the National Natural Science Foundation of China (No.11361033) and the Natural Science Foundation of Gansu Province (No.1212RJZA029).}}

\author{Pengli Lu\thanks{Corresponding author. E-mail addresses: lupengli88@163.com (\textbf{P. Lu}), dongchen199508@163.com (\textbf{C. Dong}).} \;and\; Chen Dong
\\
\footnotesize{School of Computer and Communication, Lanzhou University of Technology, Lanzhou, 730050, Gansu, P.R. China}}

\date{}

\begin{document}

\openup 0.5\jot
\maketitle

\begin{abstract}
The safety and robustness of the network have attracted the attention of people from all walks of life, and the damage of several key nodes will lead to extremely serious consequences. In this paper, we proposed the clustering H-index mixing (CHM) centrality based on the H-index of the node itself and the relative distance of its neighbors. Starting from the node itself and combining with the topology around the node, the importance of the node and its spreading capability were determined. In order to evaluate the performance of the proposed method, we use Susceptible-Infected-Recovered (SIR) model, monotonicity and resolution as the evaluation standard of experiment. Experimental results in artificial networks and real-world networks show that CHM centrality has excellent performance in identifying node importance and its spreading capability.
\bigskip


\bigskip

\end{abstract}

\section{Introduction}
In recent years, the topological structure of complex networks and the communication performance of nodes in the network have become the hot issues in current research \cite{1,2}. The most effective way to solve this problem is to abstract the communication process in the network into the communication phenomenon such as daily life, including marketing, management, education, entertainment, and so on \cite{3}. Since the communication process is far more complex than the researchers' assumption, it is still a difficult task to establish an appropriate evaluation model to determine the spreading capability of nodes in the network.

Under the unremitting efforts of the researchers, some classical theories have been proposed, including degree centrality \cite{4} that only consider the topology of nodes themselves, betweenness centrality \cite{5} that indicate whether a given node locates in shortest paths between other nodes, closeness centrality \cite{6} of a node is defined as the sum of the shortest paths from a given node to each of the others in the whole network. In addition, researchers also try to determine the influence of nodes by decomposing networks. K-core decomposing centrality \cite{7} and H-index centrality \cite{8} are widely used in various networks. However, these old methods all have their own defects. Nodes of different importance have the same level under the measure, which cannot ensure that the performance of all nodes is accurately expressed \cite{9,10,11}. On the basis of these theories, some extended measures are proposed for determination. In order to express node importance more precisely, researchers put more emphasis on the topological structures of researching nodes and their neighbors, including extended coreness centrality \cite{12}, gravity centrality \cite{13} and improved gravity centrality \cite{14}. Meanwhile, in order to make the experimental results more practical, using the spreading model is also a very common way to determine the spreading capacity of nodes \cite{15,16,17,18}. Susceptible-Infectious-Recovered (SIR) model \cite{18,19,20} is one method that has been applied most frequently in these spreading models. Multiple experimental values are obtained through a large number of experiments and the average value of these experimental values is calculated to represent the spreading capacity of each node. In the meantime, with the continuous expansion of Internet and communication operator technology in real society, social networks composed by thousands of nodes and edges have been emerged \cite{21,22}. Because of its large time complexity, these algorithms take a lot of time to get the correct capability of nodes. The existence of these networks brings some trouble to the use of these existing measures \cite{23,24,25}. In a word, it is still a problem worth studying to put forward an efficient and reasonable ranking measure to find out the important nodes in the network.

The existing methods only consider the nature of the node itself and ignore the influence of its neighbors, so they cannot accurately express the nature of node. In this paper, in order to break the limitation of existing methods, we propose a method clustering H-index mixing centrality (CHM) which based on nodes' topological structure and relative distance between nodes. Many factors play a decisive role in the importance of nodes, considering only one influencing factor and ignoring other aspects cannot get correct experimental results. Based on these theories and previous experimental results, we take the following factors into account: nodes' degree value, H-index value, local clustering coefficient, and contact distance from a given node to the other nodes in networks. In order to determine the effect of our method, we use the SIR model and kendall coefficient to compare the existing method and the proposed method. The experimental results of multiple experiments in real networks and artificial networks show that CHM method has a good performance in distinguishing the spreading capability of nodes.


\section{Method}\label{Se2}
In this paper, we denote social network as a pair $G = (V,E)$, where $V$ is the set of nodes, and $E$ is the set of edges, which represents the relations between nodes. Nodes at either end of an edge are defined as neighbors, the degree of node $v_i$ is defined as $d_i$, which represents the number of neighboring nodes of a node, and the set of these neighbors is described as $N_{i}$.

In recent years, as a new method to determine the importance of nodes, H-index centrality has been widely used but also exposed some problems. Even considering the performance of neighbor nodes, it cannot completely reflect the spreading capability of nodes themselves, and nodes of different levels often have the same H-index value. In this experiment, we take the relative distance between nodes as part of the evaluation standard, and combine degree centrality and H-index centrality to evaluate the spreading capability of nodes. Considering the scale of existing network, only neighbors whose distance is less than or equal to $3$ are calculated in this experiment (assume that the distance of the direct neighbor node of the node is $1$, the distance of the 2-order neighbor is $2$, and so on). Meanwhile, in order to improve the accuracy of H-index in evaluating the influence of nodes, we consider the sum of H-index of nodes themselves and their neighbors. The formula is as follows.
    \begin{equation}\label{adjmatix}
    \begin{aligned}
    \begin{split}
    A(i)=\left (H_{index}(v_{i})+ \displaystyle\sum_{v \in N_{i}} H_{index}(v)+ H_{index}(u_{j})+ \displaystyle\sum_{u \in N_{j}} H_{index}(u) \right)+ (\mu \times d(j))
    \end{split}
    \end{aligned}
    \end{equation}
where $j$ is the neighbor of node $i$, $N_{i}$, $N_{j}$ is the set of neighbors of node $v_{i}$, $v_{j}$, respectively. $H_{index}(v_{i})$ and $H_{index}(v_{j})$ represent the H-index value of node $v_{i}$ and $v_{j}$, $\mu$ is a tunable parameter in range $(0,1)$, $d(j)$ means the degree value of node $v_{j}$, ${d_{ij}}$ is the shortest distance between two nodes $v_{i}$ and $v_{j}$. After the verification of a large number of experiments, we assume parameter $\mu=0.5$.

The proposed method comprehensively considers the topological structure of nodes and the properties of their neighbors, and subdivides the spreading capability of each node according to the location of the neighbor nodes, so as to reflect the different importance of nodes more accurately. The clustered H-index mixing method (CHM) of node $v_{i}$ is calculated using $Eq.(2.2)$
    \begin{equation}\label{adjmatix}
    \begin{aligned}
    \begin{split}
    CHM(v_{i})=(1+C_{i}) \left (\displaystyle\sum_{j \in \varphi_{i}} \frac{A_{i}}{d^{2}_{ij}} \right)
    \end{split}
    \end{aligned}
    \end{equation}
where $C_{i}$ represents the value of clustering coefficient of node $v_{i}$. In addition, $\varphi_{i}$ represents the set of neighboring nodes with the shortest path length less than the specified length (i.e., $d_{ij} \leq r$, without loss of generality, we set $r = 3$ since the network size is so large that the computing cost is very high if $r > 3$), and node $j$ is an element of $\varphi_{i}$.

\begin{figure}
  \center
  \includegraphics[width=8cm]{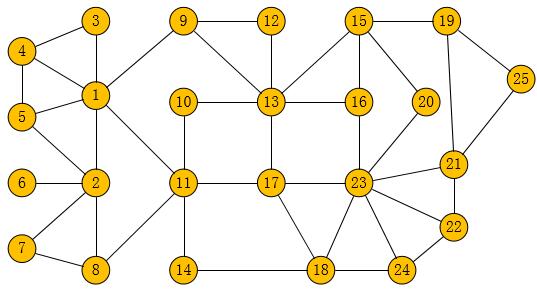}\\
  \caption{Schematic diagram}\label{1}
\end{figure}

\begin{table}
Table 1: The values of all measures on random network.

\scriptsize
\begin{tabular}{ccccccc}
 \toprule
 node numbering & Degree centrality & Betweenness centrality & Closeness centrality & K-core & H-index & CHM \\
 \midrule
    1 & 6 & 177.1667 & 0.4068 & 2 & 3 & $\mathbf{281.2431}$\\
    2 & 5 & 67.1667 & 0.3200 & 2 & 3 & 198.6611\\
    3 & 2 & 0 & 0.2963 & 2 & 2 & 77.8470\\
    4 & 3 & 1 & 0.3000 & 2 & 2 & $\mathbf{114.7529}$\\
    5 & 3 & 3.6667 & 0.3117 & 2 & 3 & 127.5879\\
    6 & 1 & 0 & 0.2449 & 1 & 3 & 39.4719\\
    7 & 2 & 0 & 0.2791 & 2 & 1 & 74.8056\\
    8 & 3 & 48.3333 & 0.3380 & 2 & 2 & 119.8969\\
    9 & 3 & 77.1667 & 0.3934 & 2 & 2 & 129.7081\\
    10 & 2 & 11.7333 & 0.3692 & 2 & 2 & 95.1003\\
    11 & 5 & 181.9667 & 0.4364 & 2 & 2 & 230.8690\\
    12 & 2 & 0 & 0.3380 & 2 & 3 & 83.3522\\
    13 & 6 & 135.5333 & 0.4286 & 2 & 2 & 289.0259\\
    14 & 2 & 16.0000 & 0.3582 & 2 & 3 & $\mathbf{89.4567}$\\
    15 & 4 & 61.3333 & 0.3582 & 2 & 2 & $\mathbf{172.1369}$\\
    16 & 3 & 15.9000 & 0.3692 & 2 & 3 & 154.5837\\
    17 & 4 & 142.3000 & 0.4528 & 2 & 4 & 223.1043\\
    18 & 4 & 35.3667 & 0.3692 & 2 & 3 & 199.8074\\
    19 & 3 & 22.5000 & 0.3000 & 2 & 2 & $\mathbf{110.8902}$\\
    20 & 2 & 3.7333 & 0.3200 & 2 & 2 & 96.4802\\
    21 & 4 & 40.1667 & 0.3158 & 2 & 3 & 173.1325\\
    22 & 3 & 3 & 0.3077 & 2 & 3 & 147.2226\\
    23 & 7 & 145.9667 & 0.4068 & 2 & 3 & $\mathbf{398.5362}$\\
    24 & 3 & 2 & 0.3038 & 2 & 3 & 147.6389\\
    25 & 2 & 0 & 0.2581 & 2 & 2 & 71.5810\\
 \bottomrule
\end{tabular}
\end{table}

In order to further compare the proposed method with the classical measures, we use the simple network as examples \cite{26,27,28}. In Figure. 1, we show the node schematic diagram of a random network, and evaluate the importance of the nodes in this network through several classical measures and the proposed method. The results are shown in Table 1. Through the Table 1 we can see that the proposed method has better effect in distinguishing node's importance, which degree centrality can divide nodes into $7$ levels, betweenness centrality is $21$ grades, closeness centrality is $18$ ranks, k-core centrality is $2$ scales, H-index is $4$ levels, however, the proposed method CHM centrality can divide all nodes into $25$ ranks.


\section{Evaluation Criterion}\label{Se3}
\subsection{Spreading model}
Spreading models can be used to evaluate the spreading capability of different nodes in the network. One of the most commonly used models is the epidemic model. The epidemic model which called Susceptible-Infected-Recovered (SIR) has been widely used in simulating the spreading process of nodes, the spreading capability of nodes is ranked through multiple experiments. In the standard stochastic SIR model, each node can be divided into three different states: Susceptible (S), Infected (I), and Recovered (R). At the beginning of a spreading process, only one node will be set to the infected state while all other nodes will be set to the susceptible state. Then, the infected nodes will spread to all the susceptible nodes connected with it in probability $\beta$. The infected nodes will recover with probability $\alpha$ and be defined as the recovery state. At the end of the whole process, there will only be two states in the whole network: susceptible nodes and recovered nodes. We only need to calculate the number of recovered nodes and record this number as the spreading capacity of the infective nodes.

After recording the spreading capability of all nodes in the network, we use kendall correlation coefficient \cite{29} to measure the performance of different centrality measures. Kendall's correlation coefficient is a measure for ranking: the similarity of data ranking list is determined by the value of different centrality measures. Intuitively, if the two variables have a similar rank, the kendall correlation coefficient is high. If the two variables have a dissimilar rank, its kendall correlation coefficient will also be low. Given any specific centrality methods, a specific kendall coefficient value can be obtained by comparing the ranking list of these methods with the ranking list calculated by SIR model, so as to evaluate the performance of this method more conveniently. The ranking list determined by one of this centrality measures is more similar to the nodes' real spreading capability if $\tau$ is more close to 1.

The ranking list $A$ denotes the ranking result of a certain centrality method, and $B$ represents the real infected numbers of all nodes in the network, let $(a_{i}, b_{i})$ be a set of random observations from two ranking data $A$ and $B$ respectively.

For any set of sequences $(a_{i}, b_{i})$ and $(a_{j}, b_{j})$, if both $a_{i} > a_{j}$ and $b_{i} > b_{j}$ or if both $a_{i} < a_{j}$ and $b_{i} < b_{j}$, they are called to be concordant. If $a_{i} > a_{j}$ and $b_{i} < b_{j}$ or if $a_{i} < a_{j}$ and $b_{i} > b_{j}$, they are called to be discordant. If $a_{i} = a_{j}$ or $b_{i} = b_{j}$, they are not taken into account. The definition of kendall's coefficient $\tau$ is as follows:
    \begin{equation}\label{adjmatix}
    \begin{aligned}
    \begin{split}
    \tau = \frac{2(x_{a}-x_{b})}{x(x-1)}
    \end{split}
    \end{aligned}
    \end{equation}
where, $x_{a}$ denotes the number of concordant pairs and $x_{b}$ denotes the number of discordant pairs, $x$ denotes the number of all pairs.

\subsection{Monotonicity}
In order to better distinguish the spreading capability of different nodes and calculate the ability of uniformly distributed nodes under the same criteria level, researchers put forward the concept of monotonicity \cite{30}, which is one of the references for evaluating the ranking methods of node influence in social networks. Monotonicity is defined as follows:
    \begin{equation}\label{adjmatix}
    \begin{aligned}
    \begin{split}
    M(A)=\left (1-\frac{\sum_{a \in A} x_{a} \times (x_{a}-1)}{B \times (B-1)}\right)^{2}
    \end{split}
    \end{aligned}
    \end{equation}
Where $A$ is the ranking list of a centrality measure, $B$ is the number of nodes of the network, $a$ is an element of list $A$, $x_{a}$ is the number of nodes in the same ranking list $a$. The range of $M(A)$ between $0$ and $1$, when the value of M(A) is $0$, all nodes have the same rank; when the value of $M(A)$ is $1$, every node assigns to unique rank in the network.

\subsection{Resolution}
Resolution is an evaluation method that reflects the different centrality measures in distinguishing the spreading performance of nodes in the network. In this paper, we examine the resolution of the method in two different ways: the cumulative distribution function (CDF) \cite{31} and the $\Delta$ \cite{32} function. The random variable $X$ represents the ranking list generated by a centrality measure, while CDF of $X$ represents the probability that the value of list $X$ is less than or equal to the given number \cite{33}. In the other word, if the CDF curve of a measure rises more slowly, its resolution will be higher. For $\Delta$ method, it can not only reflect the identification of the influence of the measure on nodes, but also represent its inner even degree. The $f(r_{A})$ in $\Delta$ function is defined as follows:
    \begin{equation}\label{adjmatix}
    \begin{aligned}
    \begin{split}
    f (r_{A}) = 1 - \frac{ \sum_{i=1}^R N^{2}_{i}}{R \cdot N^{2}}
    \end{split}
    \end{aligned}
    \end{equation}
    \begin{equation}\label{adjmatix}
    \begin{aligned}
    \begin{split}
    \Delta (A)= 1 - f(r_{A})
    \end{split}
    \end{aligned}
    \end{equation}
where, $r_{A}$ denotes the ranking list generated by centrality measure $A$. $R$ is the number of elements that do not repeat in an ordered sequence  $r_{A}$. $N_{i}$ is the number of nodes in element $i$ of $r_{A}$. $N$ represents the number of nodes in network.


\section{Introduction of network datasets and experimental results}
\subsection{Network datasets}

In this paper, a total of eight real network datasets and two artificial network datasets were involved. Among them, the real network datasets including Zachary's karate club members network (Karate) \cite{34}, Lusseau's Bottlenose Dolphins social network (Dolphins) \cite{35}, the network of selling political books about the presidential election in Amazon during 2004 (Polbooks) \cite{36}, the network of American football games schedule (Football) \cite{37}, the network of different Jazz musicians relationships (Jazz) \cite{38}, USair transportation network (USair) \cite{39}, the network of exchanging e-mail messages between members in the University Rovira Virgili (Email) \cite{40}, a social network which represents protein-protein interaction (Yeast) \cite{41}. In artificial network datasets, including Small-World network (WS) \cite{42} and Lancichinetti-Fortunato-Radicchi network (LFR) \cite{43}, both sets of artificial network datasets are generated by software Gephi. The specific data of the network used in this experiment are shown in Table 2.

\begin{table}
Table 2: Properties of the used datasets.

\begin{tabular}{cccccc}
 \toprule
 Network & ~~~~|V| & ~~~~|E| & ~~Average number & ~~Maximum degree & ~~Assortativity\\
 \midrule
 Karate & ~~~~34 &~~~~ 78 &~~ 4.588 &~~ 17 &~~ -0.4756 \\
 Dolphins &~~~~ 62 &~~~~ 159 &~~ 5.129 &~~ 12 &~~ -0.0436\\
 Polbooks &~~~~ 105 &~~~~ 441 &~~ 8.400 &~~ 25 &~~ -0.1279\\
 Football &~~~~ 115 &~~~~ 613 &~~ 10.661 &~~ 12 &~~ 0.1624\\
 Jazz &~~~~ 198 &~~~~ 2742 &~~ 27.967 &~~ 100 &~~ 0.0202\\
 USair &~~~~ 332 &~~~~ 2126 &~~ 12.81 &~~ 139 &~~ -0.2079\\
 Email &~~~~ 1133 &~~~~ 5451 &~~ 9.622 &~~ 71 &~~ 0.0782\\
 WS &~~~~ 2000 &~~~~ 6012 &~~ 6.021 &~~ 11 &~~ -0.0563\\
 LFR-2000 &~~~~ 2000 &~~~~ 4997 &~~ 9.988 &~~ 39 &~~ -0.0032\\
 Yeast &~~~~ 2361 &~~~~ 7181 &~~ 6.083 &~~ 65 &~~ -0.0489\\
 \bottomrule
\end{tabular}
\end{table}

\subsection{Result of spreading model}
In the first experiment, in order to determine the accuracy of the proposed method in identifying the spreading capability of nodes compared with existing methods, kendall coefficient $\tau$ calculated by SIR model was used to verify it. The value of kendall coefficient is between $0$ and $1$. The closer $\tau$ is to $1$, the higher the similarity between the two sets of ranking lists. $\beta$ and $\beta_{th}$ of each network dataset under the SIR process are shown in Table 3, which also shows the results obtained through experiments.

The method introduced in section 3.1 is used to evaluate the accuracy of the ranking measure in artificial networks and real-world networks respectively. In this experiment, we consider that if the $\beta$ value is too large, the spreading capacity and topological structure of the origination node are no longer important to the whole SIR process, so we choose a relatively small $\beta$ value to ensure the rationality of the experiment. In other words, the experimental results do not change much depending on the initial node selected. After $100$ repeated experiments, we calculated the average of infected nodes at the end of these experimental processes as the spreading capacity of the nodes.

It can be seen from Table 3 that the proposed method CHM shows strong competitiveness in most networks and can effectively reflect the spreading process of real datasets. Only in Football network, due to network characteristics, the accuracy of CHM method is not as good as the traditional betweenness centrality.

In order to further reflect the accuracy of the proposed method, we selected different value of $\beta$ to calculate the $\tau$ values under various networks, and drew a graph as shown in Figure. 2. The cruve in Figure. 2 reflects the $\tau$ values of various measures under different $\beta$ values. From the curve trend in the graph, we can see that the proposed method is closer to reality than existing methods and the measure can improve the accuracy. With the continuous expansion of $\beta$ value, the accuracy of CHM centrality gradually improves. Especially under the USair and Yeast network, CHM centrality on the basis of the existing methods have greatly improved the accuracy.

\begin{table}
Table 3: The kendall $\tau$ value of each method in 10 networks with a given $\beta$ value.

 \begin{tabular}{ccccccccc}
 \toprule
 Network & $\beta_{th}$ & $\beta$ & $\tau$(DC, $\sigma$) & $\tau$(BC,$\sigma$) & $\tau$(CC,$\sigma$) & $\tau$(KS,$\sigma$)
 & $\tau$(H,$\sigma$) & $\tau$(CHM,$\sigma$)\\
 \midrule
 Karate & 0.129 & 0.17 & 0.7035 & 0.6345 & 0.7615 & 0.6522 & 0.6669 & $\mathbf{0.8595}$\\
 Dolphins & 0.147 & 0.16 & 0.8029 & 0.5615 & 0.6083 & 0.7266 & 0.8581 & $\mathbf{0.8655}$\\
 Polbooks & 0.0838 & 0.16 & 0.7554 & 0.3794 & 0.4084 & 0.7224 & 0.7943 & $\mathbf{0.8559}$\\
 Football & 0.0932 & 0.20 & $\mathbf{0.5034}$ & 0.3021 & 0.4258 & 0.1319 & 0.4595 & 0.4789\\
 Jazz & 0.026 & 0.12 & 0.8175 & 0.5032 & 0.7778 & 0.7475 & 0.8309 & $\mathbf{0.8786}$\\
 USair & 0.0225 & 0.07 & 0.7448 & 0.5628 & 0.8119 & 0.7721 & 0.7708 & $\mathbf{0.8935}$\\
 Email & 0.0535 & 0.10 & 0.7979 & 0.6504 & 0.8209 & 0.8077 & 0.8217 & $\mathbf{0.8936}$\\
 WS & 0.1559 & 0.22 & 0.6569 & 0.6127 & 0.5985 & 0.1229 & 0.5175 & $\mathbf{0.6481}$\\
 LFR-2000 & 0.0477 & 0.18 & 0.8054 & 0.7624 & 0.8125 & 0.4236 & 0.7622 & $\mathbf{0.8121}$\\
 Yeast & 0.0600 & 0.15 & 0.6692 & 0.5943 & 0.6271 & 0.7087 & 0.7120 & $\mathbf{0.8780}$\\
 \bottomrule
\end{tabular}
\end{table}

\begin{table}
Table 4: The M value of ranking list generated by different measures in 10 networks.

\begin{tabular}{ccccccc}
 \toprule
 Network & ~~~~M(DC) & ~~~~M(BC) & ~~~~M(CC) & ~~~~M(KS) & ~~~~M(H) & ~~~~M(CHM)\\
 \midrule
 Karate & ~~~~0.7079 & ~~~~0.7754 & ~~~~0.8993 & ~~~~0.4958 & ~~~~0.5766 & ~~~~$\mathbf{0.9577}$ \\
 Dolphins & ~~~~0.8312 & ~~~~0.9623 & ~~~~0.9737 & ~~~~0.3769 & ~~~~0.6841 & ~~~~$\mathbf{0.9979}$ \\
 Polbooks & ~~~~0.8252 & ~~~~0.9974 & ~~~~0.9847 & ~~~~0.4949 & ~~~~0.7067 & ~~~~$\mathbf{0.9999}$ \\
 Football & ~~~~0.3637 & ~~~~$\mathbf{0.9999}$ & ~~~~0.9488 & ~~~~0.0003 & ~~~~0.2349 & ~~~~0.9994 \\
 Jazz & ~~~~0.9659 & ~~~~0.9885 & ~~~~0.9878 & ~~~~0.7944 & ~~~~0.9583 & ~~~~$\mathbf{0.9996}$ \\
 USair & ~~~~0.8586 & ~~~~0.6970 & ~~~~0.9892 & ~~~~0.8114 & ~~~~0.8355 & ~~~~$\mathbf{0.9952}$ \\
 Email & ~~~~0.8875 & ~~~~0.9400 & ~~~~0.9988 & ~~~~0.8089 & ~~~~0.8584 & ~~~~$\mathbf{0.9999}$ \\
 WS & ~~~~0.5922 & ~~~~$\mathbf{0.9999}$ & ~~~~0.9987 & ~~~~0.0002 & ~~~~0.2904 & ~~~~$\mathbf{0.9999}$ \\
 LFR-2000 & ~~~~0.8760 & ~~~~$\mathbf{0.9999}$ & ~~~~0.9951 & ~~~~0.0385 & ~~~~0.7184 & ~~~~$\mathbf{0.9999}$ \\
 Yeast & ~~~~0.7210 & ~~~~0.7012 & ~~~~0.9964 & ~~~~0.6643 & ~~~~0.6873 & ~~~~$\mathbf{0.9965}$ \\
 \bottomrule
\end{tabular}
\end{table}

\begin{figure}
  \center
  \includegraphics[width=16cm]{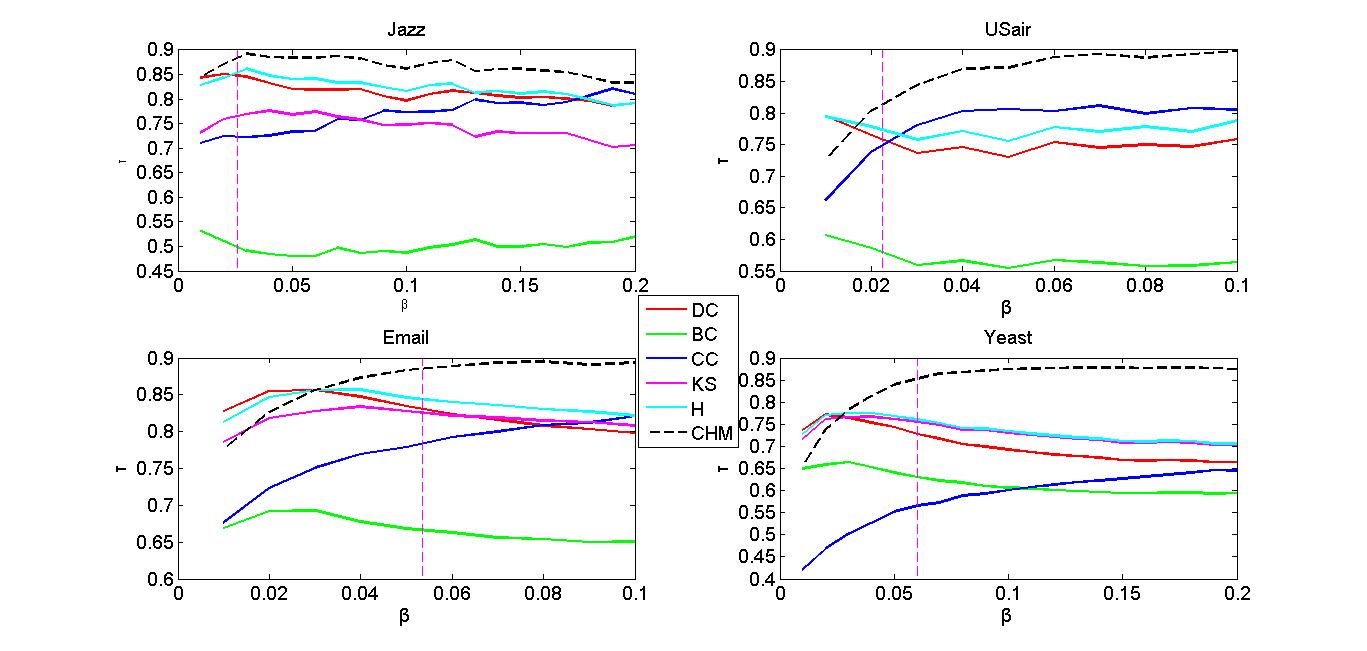}\\
  \caption{Kendall $\tau$ value curve under different infection and recovery rates on Jazz, USair, Email, Yeast networks.}\label{*}
\end{figure}

\begin{figure}
  \center
  \includegraphics[width=16cm]{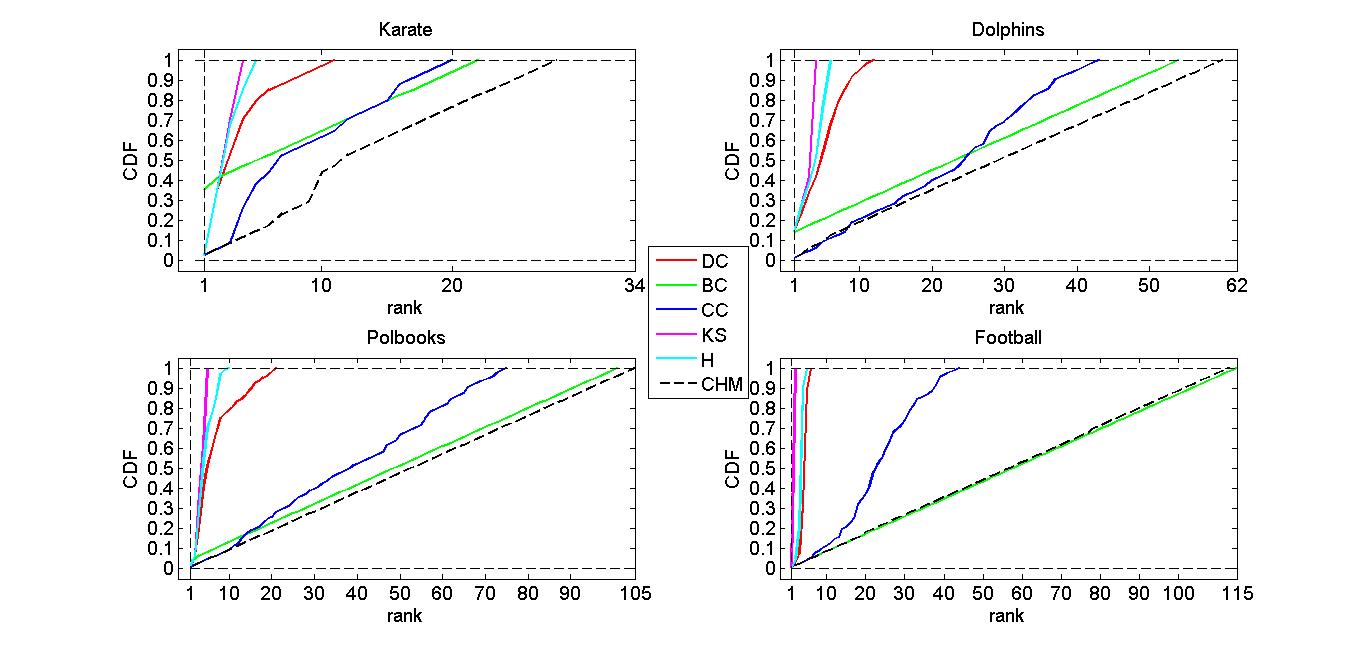}\\
  \caption{The CDF curve of all measures on Karate, Dolphins, Polbooks, Football networks.}\label{*}
\end{figure}

\begin{figure}
  \center
  \includegraphics[width=16cm]{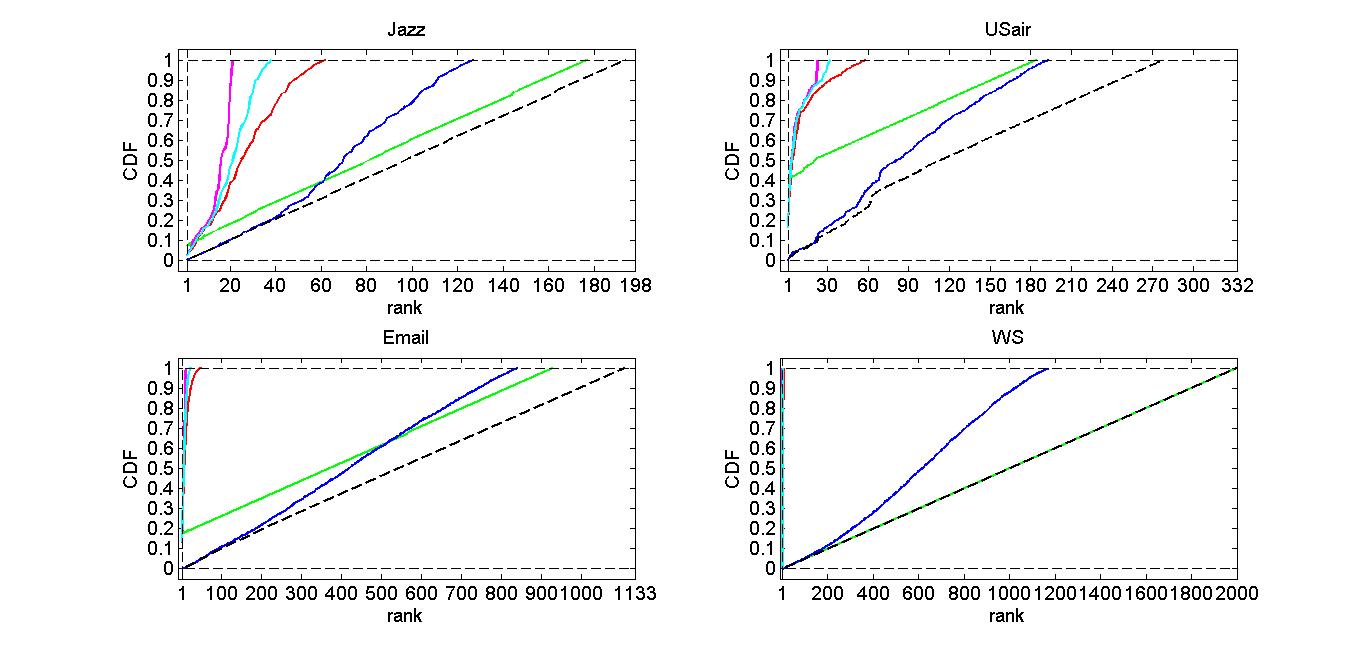}\\
  \caption{The CDF curve of all measures on Jazz, USair, Email, WS networks.}\label{*}
\end{figure}

\begin{figure}
  \center
  \includegraphics[width=16cm]{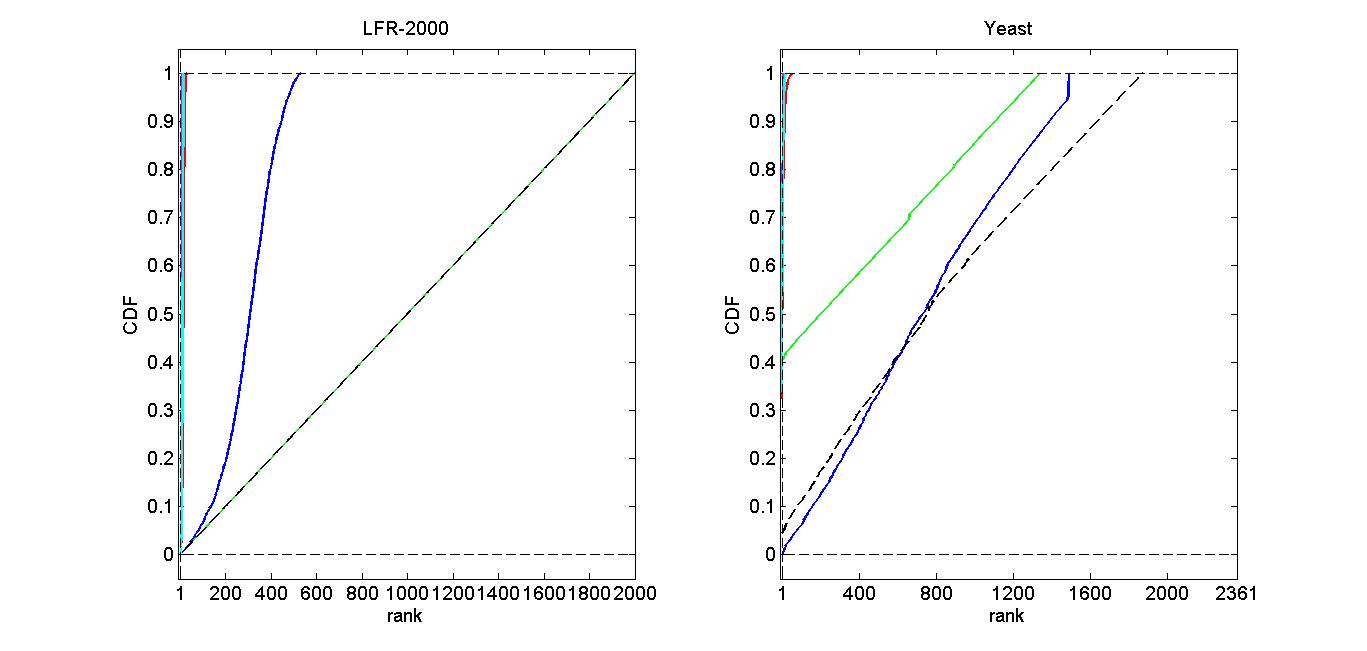}\\
  \caption{The CDF curve of all measures on LFR-2000, Yeast networks.}\label{*}
\end{figure}

\subsection{Result of monotonicity}
In the second experiment, we used monotonicity as a standard for various measures. Table 4 shows the monotonic values of various measures on different networks. The higher the value of monotonicity, the stronger the ability of this measure to distinguish the importance of nodes.

In Table 4, due to different network properties, the monotonicity values of various measures in different networks are not consistent, and the values
of the proposed methods are close to $1$ in all networks, showing strong competitiveness. In Football, WS and LFR-2000 networks, betweenness centrality also shows strong node discrimination capability, while the proposed method still has some advantages over betweenness centrality.

\begin{table}
Table 5: The $\Delta$ value of different measures on 10 networks.

\normalsize
\begin{tabular}{ccccccc}
 \toprule
 Network & ~~~~$\Delta$ DC & ~~~~$\Delta$ BC & ~~~~$\Delta$ CC & ~~~~$\Delta$ KS & ~~~~$\Delta$ H & ~~~~$\Delta$ CHM \\
 \midrule
 Karate & ~~~~$\frac{(4)}{1.67e-02}$ & ~~~~$\frac{(3)}{6.60e-03}$ & ~~~~$\frac{(2)}{4.00e-03}$ & ~~~~$\frac{(6)}{7.92e-02}$ & ~~~~$\frac{(5)}{5.26e-02}$ & ~~~~$\frac{(1)}{1.80e-03}$\\
 Dolphins & ~~~~$\frac{(4)}{8.60e-03}$ & ~~~~$\frac{(2)}{6.46e-04}$ & ~~~~$\frac{(3)}{6.78e-04}$ & ~~~~$\frac{(6)}{9.90e-02}$ & ~~~~$\frac{(5)}{3.10e-02}$ & ~~~~$\frac{(1)}{2.86e-04}$\\
 Polbooks & ~~~~$\frac{(4)}{4.80e-03}$ & ~~~~$\frac{(2)}{1.07e-04}$ & ~~~~$\frac{(3)}{2.29e-04}$ & ~~~~$\frac{(6)}{6.06e-02}$ & ~~~~$\frac{(5)}{1.67e-02}$ & ~~~~$\frac{(1)}{9.07e-05}$\\
 Football & ~~~~$\frac{(4)}{6.70e-02}$ & ~~~~$\frac{(1)}{7.56e-05}$ & ~~~~$\frac{(3)}{7.82e-04}$ & ~~~~$\frac{(6)}{4.91e-01}$ & ~~~~$\frac{(5)}{1.04e-01}$ & ~~~~$\frac{(2)}{7.96e-05}$\\
 Jazz & ~~~~$\frac{(4)}{3.75e-04}$ & ~~~~$\frac{(2)}{6.08e-05}$ & ~~~~$\frac{(3)}{8.76e-05}$ & ~~~~$\frac{(6)}{5.40e-03}$ & ~~~~$\frac{(5)}{9.53e-04}$ & ~~~~$\frac{(1)}{2.71e-05}$\\
 USair & ~~~~$\frac{(4)}{1.30e-03}$ & ~~~~$\frac{(3)}{9.11e-04}$ & ~~~~$\frac{(2)}{4.35e-05}$ & ~~~~$\frac{(6)}{4.40e-03}$ & ~~~~$\frac{(5)}{2.80e-03}$ & ~~~~$\frac{(1)}{1.95e-05}$\\
 Email & ~~~~$\frac{(4)}{1.20e-03}$ & ~~~~$\frac{(3)}{3.38e-05}$ & ~~~~$\frac{(2)}{1.77e-06}$ & ~~~~$\frac{(6)}{9.20e-03}$ & ~~~~$\frac{(5)}{3.10e-03}$ & ~~~~$\frac{(1)}{8.42e-07}$\\
 WS & ~~~~$\frac{(4)}{2.56e-02}$ & ~~~~$\frac{(2)}{2.51e-07}$ & ~~~~$\frac{(3)}{9.73e-07}$ & ~~~~$\frac{(6)}{4.92e-01}$ & ~~~~$\frac{(5)}{9.23e-02}$ & ~~~~$\frac{(1)}{2.51e-07}$\\
 LFR-2000 & ~~~~$\frac{(4)}{2.20e-03}$ & ~~~~$\frac{(1)}{2.50e-07}$ & ~~~~$\frac{(3)}{5.54e-06}$ & ~~~~$\frac{(6)}{8.93e-02}$ & ~~~~$\frac{(5)}{9.60e-03}$ & ~~~~$\frac{(1)}{2.50e-07}$\\
 Yeast & ~~~~$\frac{(4)}{2.80e-03}$ & ~~~~$\frac{(3)}{1.21e-04}$ & ~~~~$\frac{(2)}{1.49e-06}$ & ~~~~$\frac{(6)}{1.85e-02}$ & ~~~~$\frac{(5)}{8.60e-03}$ & ~~~~$\frac{(1)}{1.15e-06}$\\
 \bottomrule
\end{tabular}
\end{table}

\subsection{Result of resolution}
In the third experiment, two different methods are used to evaluate the resolution of nodes by various centrality measures. The details are shown in the Figure. 3 - Figure. 5 and Table 5 respectively.

Figure. 3 - Figure. 5 show the CDFs curve of ranking sequences in different networks. The inclination degree of the curve in the figure can be used to reflect the dispersion degree of the ranking list. The slower the curve rises, the more discrete the list is. It can be seen from the figure that in $10$
network datasets, the proposed method CHM has good performance. In Football network, betweenness centrality is better than CHM and CHM is better than other measures; in WS network and LFR-2000 network, CHM centrality has the same excellent performance as betweenness centrality and better than other measures; while in the remaining $7$ networks, CHM centrality is better than betweenness centrality and other measures. In Table 5, the proposed method CHM's values are all smaller than those of other methods. In both artificial and real-world networks, the proposed method has great effect in distinguishing the importance of nodes. Only in Football network, the value of betweenness centrality method is greater than that of CHM.


\section{Conclusion}
With the sustainable growth of the users in social networks and the frequency of exchanging information between users, it becomes an urgent task to build a network framework that can transmit information in real time and efficiently. For researchers, finding as few key nodes as possible to maximize the influence on the whole communication process has become the most important thing. Many measures have been used to identify important nodes and calculate their spreading capacity. However, these existing measures are all flawed and have not been able to meet the researchers' expectations. In this paper, by considering the topology of the node itself and the properties of its neighbors, the performance of the node can be judged more accurately. The experimental results show that the proposed method is effective in determining the spreading capacity and importance of nodes in both artificial and real-world networks. Clustering H-index mixing measure can be investigated in both directed networks and undirected networks in future research.

\medskip


\end{document}